\documentclass[%
 reprint,
 amsmath,amssymb,
 aps,
]{revtex4-2}

\usepackage{graphicx}
\usepackage{dcolumn}
\usepackage{bm}
\usepackage{tikz}

\begin{document}

\preprint{APS/123-QED}

\title{Gravitational Lensing in a Universe with matter and Cosmological Constant}

\author{Pedro Bessa}
\affiliation{PPGCosmo - UFES,\\
Universidade Federal do Espírito Santo,\\
Vitória, 29075-910, Brazil}
\email{pedvbessa@gmail.com}
\author{Oliver F Piattella}%
 \email{of.piattella@uninsubria.it}
\affiliation{Dipartimento di Scienza e Alta Tecnologia,\\ Universit\`a degli Studi dell'Insubria e INFN,\\ via Valleggio 11, I-22100 Como (Italy)}
\altaffiliation[Also at ]{Cosmo-ufes, Universidade Federal do Esp\'irito Santo, avenida F. Ferrari 514, 29075-910 Vit\'oria, Esp\'irito Santo (Brazil)}
\begin{abstract}
We extend the results obtained in \cite{Piattella_2016, mcvittie_2015} and \cite{Park_2008} for gravitational lensing in the McVittie metric by including the effect of the transition from the matter-dominated epoch of the Universe to the $\Lambda$-dominated era. We derive a formula that agrees with the previous results for the McVittie metric at lowest order, and compare the lensing angle predictions obtained from the Schwarzschild approximation, the McVittie model and higher order corrections to the McVittie model. In doing this, we test if, beyond the correction from the accelerated expansion of the Universe, there is a need for including the matter content of the Universe in modeling lens systems at the redshifts observed in lens systems.
We investigate if there is a need for a modification of the lens equation from these corrections, and if so, to which order and whether it is measurable.
We find that while the effect is of the same order as the one calculated previously, there is no significant contribution to the bending angle, as the 1st order effect is already of order $\mathcal{O}(\theta_O^4)$ in the observed angle.
\end{abstract}
\date{\today}

\maketitle

\section{Introduction}
The matter of whether the cosmological constant $\Lambda$ has a significant impact on gravitational lensing and how to measure its effects has been a source of debate in the literature \cite{Rindler_2007,Ishak_2010, Aghili_2016, Tian_2017, Hu_2022,Piattella_2016}. The current understanding is that there is a small effect due to $\Lambda$ on the bending angle at the lens plane \cite{Rindler_2007,Park_2008,Faraoni_2017,Takizawa22}, although it is too small to be detected in the lens regimes so far observed \cite{Piattella_2016,Hu_2022}, a correction of order $\mathcal{O}(10^{-11})$ in the mass estimate of the lens  for a typical lensing system \cite{mcvittie_2015,Butcher_2016}; though see reference \cite{He_2017} for views which relate these detections to dark energy equations of state.

One of the main issues around the problem of assessing the effects of $\Lambda$ on the bending angle is how to model the lens in an expanding spacetime, and whether the spacetime metric should incorporate the effects of the expansion of the Universe instead of these effects affecting lensing only through its dependence on angular diameter distances \cite{Simpson_2010,Faraoni_2017,Ishak_2010,Virbhadra_22,Virbhadra22_2}. One of the paths to model these lenses already within an expanding spacetime is through the use of the McVittie metric \cite{McVittie:1933zz,Park_2008}.

In \cite{Piattella_2016} the McVittie metric is used to model the embedding of the lens in an expanding spacetime. It was shown that, at the 0th order the Hubble flow does not modify the expression for the bending angle. At 1st order, however, there is an effect, with contributions proportional to the redshifts of the lens and the source. The physical explanation for this is that the light rays emitted by the source are spread by the Hubbe flow, requiring thus some more convergence, hence a larger deflection angle, in order to reach the observer. In \cite{Piattella_2016} this 1st order correction is computed assuming a de Sitter space, hence a constant Hubble flow. In this paper we extend this calculation by admitting a more realistic cosmological model, in which matter is present and so in which the Hubble factor grows with the redshift.

The 1st order corrections to the bending angle (the leading order being proportional to the compactness of the lens) computed in \cite{Piattella_2016,mcvittie_2015} depend on the assumption of a constant Hubble factor $H_0$, which accounts for the current accelerated phase of expansion of the Universe due to a cosmological constant $\Lambda$ \cite{SupernovaSearchTeam:1998fmf}, and is a good approximation at low redshifts, $z \lesssim 0.3$ \cite{mcvittie_2015}. Current observations of lensing systems are capable of detecting lens-source systems where the redshifts involved are of order $\mathcal{O}(1)$ \cite{SLACS_2017}. For instance, some lenses have redshifts $z > 0.3$, which are outside the regime of validity of this approximation, as well as lenses at redshifts $z>1$ \cite{SLACS_2009}, which are well outside the $\Lambda$-dominated era, considering the transition redshift (from the matter-dominated epoch to the $\Lambda$-dominated one) being given by $z_\Lambda = (2\Omega_\Lambda/\Omega_m)^{1/3} -1 \approx 0.67$ for the approximated values $\Omega_m \approx 0.3$ and $\Omega_\Lambda \approx 0.7$.

With this in mind, we are motivated to test the impact of including the effect of matter domination on the deflection angle calculated from the McVittie metric \cite{Piattella_2016}. This is already accounted for in the angular diameter distance in the lens equation \cite{Schneider}; however, as stated previously, when one goes beyond the low redshift approximation, there are correction terms in the bending angle. 

Here we obtain an explicit, analytic formula for the angular diameter distance $D$ in a matter + $\Lambda$ universe, and invert this relation in order to calculate the full bending angle, to order $\mathcal{O}(D^3)$. We compare our results to the standard one based on the Schwarzschild metric, and to the 1st order for the McVittie metric. We also extend the results obtained in \cite{Park_2008}, where the higher order corrections were calculated for the 1st time in the McVittie metric.

At the end, we find that while the bending angle is modified by the higher order terms related to the matter dominated epoch, the corrections are of the same order as the ones previously found by including the effect of the cosmological constant, which are of small order and should not modify the standard lensing formalism, thus corroborating the current consensus on the effect of $\Lambda$ on gravitational lensing.

\section{Distances in a Universe with matter and a cosmological constant}

One can model the dynamics of the large scale expansion of the Universe with a flat Friedmann-Lema\^itre-Robertson-Walker (FLRW) spacetime:
\begin{align}
\label{flrw}
    ds^2 = -dt^2 + a^2(t)\delta_{ij}dx^i dx^j\;,
\end{align}
and a matter content given by a pressureless matter fluid and a cosmological constant $\Lambda$. The dynamics is given by the Friedmann equation:
\begin{equation}
\label{friedmann}
    H^2 = H_0[\Omega_m(1+z)^3 + \Omega_\Lambda]\;,
\end{equation}
where $H_0$ is the Hubble constant and the $\Omega$'s are the density parameters of pressureless matter and the cosmological constant. 

The McVittie metric in flat spacetime is given by

\begin{equation}
    ds^2 = \frac{(1-\mu(t))^2}{(1+\mu(t))^2}dt^2 + (1+\mu(t))^4a^2(t)\delta_{ij}dx^idx^j\;,
\end{equation}
where $\mu(t)\equiv M/a(t)\rho$; $M$ is the mass of the point-like mass-particle in the spacetime, $\rho$ is the radial coordinate and $a(t)$ is the scale factor, the same as given in \eqref{flrw}.

Following \cite{Piattella_2016}, the McVittie metric can be regarded, sufficiently far from the point-like object, as a perturbed FLRW-like metric in the Newtonian gauge: 
\begin{equation}
\label{McVittie_approx}
   ds^2 = -(1-4\mu)dt^2 + (1+4\mu)a(t)^2\gamma_{ij}dx^idx^j\;,
\end{equation}
where the gravitational potential $2\mu$ is given by $2\mu = M/a(t)\rho$, so it is not an actual perturbative degree of freedom, as in standard cosmological perturbation theory. We shall use the form \eqref{McVittie_approx} of the metric when referring to the McVittie metric throughout this paper.

\subsection{Comoving and angular-diameter distances}

Using dimensionful quantities, the cosmological distances, of the order $cH_0^{-1}$ measured for observers far from the source object surroundings, described by the metric \eqref{McVittie_approx}, should not be affected by the local effects of the mass, as we shouldn't expect a measurable gravitational interaction between source and observer at these scales. Thus, there is no issue in using the metric \eqref{flrw} in place of \eqref{McVittie_approx} for far away observers.

From this assumption, for the derivation of the bending angle, one needs the angular-diameter distances between observer $O$, source $S$ and lens $L$. In a Universe with matter and cosmological constant, using equation \eqref{friedmann}, the comoving distance between $S$ and $O$, defined as $\eta_{SO}$, is: 
\begin{eqnarray}
    \eta_{SO} = \int_{t_O}^{t_S} \frac{dt'}{a(t')} = \int_{z_S}^{z_O} \frac{dz'}{H(z')}\nonumber\\
    = \frac{1}{H_0}\int_{z_O}^{z_S} \frac{dz'}{[\Omega_m(1+z')^3 + \Omega_\Lambda]^{1/2}}\;.
\end{eqnarray}
The angular-diameter distance $D_{SO}$ between $S$ and $O$ is then given by:
\begin{equation}
\label{angdiam_dis}
    D_{SO} = \frac{1}{(1 + z_S)}\frac{1}{H_0}\int_{z_O}^{z_S} \frac{dz'}{[\Omega_m(1+z')^3 + \Omega_\Lambda]^{1/2}}\;.
\end{equation}
This integral can be written as a Gaussian hypergeometric function , inside the interval defined by $|\Omega_m/\Omega_\Lambda(1+z)^3|\leq 1$ \cite{abramowitz1964handbook}:

\begin{equation}
\label{hypergeo_form}
    H_0D_{SO} = \frac{1}{\Omega_\Lambda^{1/2}}\;{}_2F_1\left(\frac{1}{3},\frac{1}{2};\frac{4}{3};-\frac{\Omega_m }{\Omega_\Lambda}(1+z)^3\right)\Bigg|_O^S\;.
\end{equation}
Since $\mathbf{Re}(c-b-a) = 4/3 - 1/2 - 1/3 = 1/2 > 0$, one can write the hypergeometric function as the so called Gauss series \cite{abramowitz1964handbook},

and we can then finally write the comoving and angular-diameter distances as:
\begin{eqnarray}
\label{redshift_distance}
   H_0 \eta_{SO} = \frac{(1+z)}{\Omega_\Lambda^{1/2}}\frac{\Gamma(\frac{4}{3})}{\Gamma(\frac{1}{2})\Gamma(\frac{1}{3})}\times\nonumber\\
   \times\sum_{n=0}^\infty \frac{\Gamma(\frac{1}{2}+n)\Gamma(\frac{1}{3}+n)}{\Gamma(\frac{4}{3}+n)n!}\left[-\frac{\Omega_m}{\Omega_\Lambda}(1+z)^3\right]^{n}\Bigg|_O^S\;,\\
H_0 D_{SO} = \frac{1}{\Omega_\Lambda^{1/2}}\frac{\Gamma(\frac{4}{3})}{\Gamma(\frac{1}{2})\Gamma(\frac{1}{3})}\times\nonumber\\
\times\sum_{n=0}^\infty \frac{\Gamma(\frac{1}{2}+n)\Gamma(\frac{1}{3}+n)}{\Gamma(\frac{4}{3}+n)n!}\left[-\frac{\Omega_m}{\Omega_\Lambda}(1+z)^3\right]^{n}\Bigg|_O^S\;.
\end{eqnarray}
For $\Omega_m = 0$ (hence $\Omega_\Lambda = 1$) and $z_O = 0$ one can check that:
\begin{equation}
\label{first_order_redshift}
    H_0 \eta_{SO} = z_S\;,
\end{equation}
which is the result expected for a $\Lambda$-dominated (de Sitter) Universe, and the comoving distance used in \cite{mcvittie_2015}.

This general expression is valid up to values $z\approx 1.3$ of the redshift, due to the analytical properties of the series \eqref{redshift_distance}. This makes the expression valid for many of the lens systems detected so far, as one can check in the surveys \cite{SLACS_2009,SLACS_2017,BELLS_1,BELLS_2}.

\section{Lensing angle for higher redshifts}

\begin{figure}[!h]
    \centering
    \includegraphics[scale=.43]{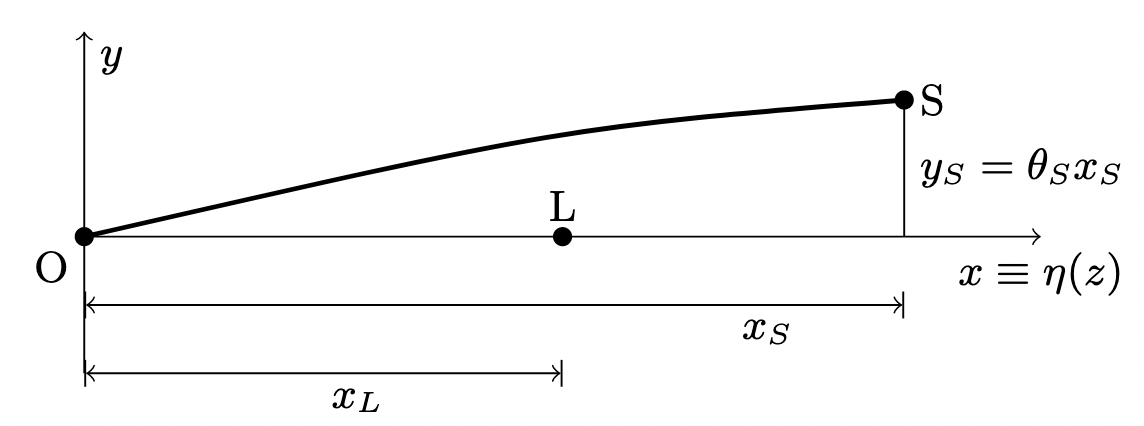}
    \caption{Lensing configuration. The comoving distance to the lens \textbf{L} is taken as a characteristic scale. The actual position of the source \textbf{S} is $y_S$.}
\label{figuretrajectory}
\end{figure}

The derivation of the bending angle for the McVittie metric follows the calculations in  \cite{Piattella_2016} and \cite{mcvittie_2015}, and the lensing configuration is illustrated in Fig.~\ref{figuretrajectory}

We define the radial distance to the lens $x_L$ as a characteristic length scale to which we normalize all the other distances. So, the equation describing the trajectory of the light ray from the source to the observer can be written as follows \cite{Piattella_2016}:
\begin{equation}
\label{lensing_dim}
    \frac{d^2 Y}{dX^2} = -\alpha \frac{Y}{a(X)[(X-1)^2 + Y^2]^{3/2}}\;,
\end{equation}
where $Y\equiv y/x_L$, $X\equiv x/x_L$ and $\alpha \equiv 2M/x_L$.

Since we keep the leading order in $\alpha$, we need just to put the zero order solution for $Y$ on the right hand side. That is, $Y = Y_S$. Note that $\alpha = \mathcal O(10^{-11})$, so the approximation employed is fair. Indeed, as we shall see, corrections due to a higher Hubble flow at large redshift are way larger than the contribution of $\alpha^2$. Recall that $y_S = \theta_S x_S$, as in figure, and this is the zero order solution.

For small angles, one can approximate $\tan\theta \approx \theta$, and from the definition of the lensing angle, which is related to the slope of the radial photons as $dy/dx \approx \tan\theta$ for small angles, equation \eqref{lensing_dim} becomes:
\begin{equation}
\label{angle_eq}
    \frac{d\theta}{dX} = -\alpha\frac{Y_S}{a(X)[(X-1)^2 + Y_S^2]^{3/2}}\;.
\end{equation}

The bending angle is then \textit{defined} as:
\begin{equation}
    \delta = \int_{X_S}^0\frac{d\theta}{dX}\;. \label{lensdef}
\end{equation}
In Ref.~\cite{Piattella_2016}, this equation is solved at $0$th order, which amounts to the standard result for the deflection by a point mass \cite{Schneider}; and in the first order approximation in $\alpha$, which gives the correction due to the embedding of the observer-lens-source system in a de Sitter space, for which $H(z) = H_0$ is constant.

The assumption of a constant Hubble factor means that we have a $\Lambda$ dominated Universe throughout the entire redshift range. This assumption breaks down for redshifts $z > 0.3$ in \cite{mcvittie_2015}, since there is significant contribution of the matter density to the cosmological distances involved in the lensing angle.

Lensing systems detected by modern surveys have lenses with redshifts $z>0.4$ and sources with redshifts $z>1$ \cite{SLACS_2017}, the former well outside the  $\Lambda$ dominated approximation; and the latter well outside the matter-$\Lambda$ transition redshift, which is constrained at $z_\Lambda\approx0.6$ both from early and late-time cosmology \cite{transition_redshift}.

This leads one to ask if the resulting correction to the lensing angle obtained, for instance, in equation (60) of \cite{Piattella_2016}, when calculated for lensing systems at higher redshifts, is significantly underestimated due to a failure of the low redshift approximation.
In the following subsection we obtain an analytic expression for $z(X)$, and integrate the resulting right hand side of \eqref{angle_eq} to obtain an expression for the bending angle including the effects of the matter phase.

\subsection{General expression for the bending angle}

In Eq.~\eqref{angle_eq} we must make the dependence $a(X)$ explicit. In order to achieve this, we first change to the redshift, since $a(X) = 1/[1 + z(X)]$, then we expand the inverse function $X(z)$ in a series of powers of $z$. Since $X(z)$ is a monotonically increasing function of the redshift, it is invertible. 

One can, then, obtain an inverse function from the coefficients of the series \eqref{redshift_distance}, or by the inverse power series of  \eqref{hypergeo_form}. This is best done through computational methods, such as using MATHEMATICA.

We truncate the inverse power series to third order, and obtain, explicitly

\begin{align}
&X(z) \equiv \frac{\eta(z)}{x_L} \longleftrightarrow z(H_0x_LX) = X^{-1}(z),\notag \\
\implies&z(X) = H_0 x_LX + a_1(H_0x_LX)^2 + a_2(H_0 x_LX)^3,&
\end{align}
where the numerical values of the coefficients $a_1$ and $a_2$ can be found in appendix A.

Using this approximation for the redshift, Eq.~\eqref{angle_eq} then becomes:
\begin{align}
\label{full_ang_eq}
    &\frac{d\theta}{dX} = \nonumber\\ 
   &-\alpha\frac{Y_S\left[1 + H_0 x_LX + a_1(H_0 x_L X )^2 + a_2(H_0 x_L X)^3\right]}{[(X-1)^2 + Y_S^2]^{3/2}}\;.
\end{align}
One can check that Eq.~\eqref{full_ang_eq} reduces at first order in $X$ to Eq.~(53) of Ref.~\cite{Piattella_2016}.

We now use the series \eqref{redshift_distance} to write $x_L$ explicitly as a function of the lens redshift $z_L$, and denote by $b_2, b_3$ the first coefficients of this expansion. Again, we keep the approximation to third order in the redshift of the lens. Our final expression for $d\theta/dX$ is then
\begin{flalign}
\label{third_order_tangent}
    &\frac{d\theta}{dX} = \notag\\
    &\frac{-\alpha Y_S}{[(X-1)^2 + Y_S^2]^{3/2}}&&\big[1+ (z_L+b_1z_L^2+b_2z_L^3)X \notag\\
    &&&+ (a_1z_L^2 + 2a_1b_1z_L^3)X^2 + a_2(z_L X)^3\big].
\end{flalign}

Integrating this equation from the origin to the source, as defined in Eq.~\eqref{lensdef}, we obtain the bending angle. Here we write the leading order contribution in $z_L$: 

\begin{widetext}
\begin{align}
    \delta^{(3)} = \alpha Y_S &\Bigg[-\frac{2 a_1 b_1 X_S \left(Y_S^2-1\right)}{Y_S^2 \sqrt{X_S^2-2 X_S+Y_S^2+1}}-\frac{2 a_1 b_1 \left(Y_S^2+1\right)}{Y_S^2 \sqrt{X_S^2-2 X_S+Y_S^2+1}}+2 a_1 b_1 \log \left(-\sqrt{X_S^2-2 X_S+Y_S^2+1}-X_S+1\right)\notag \\
    &+\frac{a_2 X_S^2}{\sqrt{X_S^2-2 X_S+Y_S^2+1}}+\frac{a_2 X_S}{Y_S^2 \sqrt{X_S^2-2 X_S+Y_S^2+1}}-\frac{5 a_2 X_S}{\sqrt{X_S^2-2 X_S+Y_S^2+1}}+\frac{2 a_2 \left(Y_S^2+1\right)}{\sqrt{X_S^2-2 X_S+Y_S^2+1}}\notag \\
    &-\frac{a_2 \left(Y_S^2+1\right)}{Y_S^2 \sqrt{X_S^2-2 X_S+Y_S^2+1}}+3 a_2 \log \left(-\sqrt{X_S^2-2 X_S+Y_S^2+1}-X_S+1\right)\notag \\
    &+\frac{b_2 X_S}{Y_S^2 \sqrt{X_S^2-2 X_S+Y_S^2+1}}-\frac{b_2 \left(Y_S^2+1\right)}{Y_S^2 \sqrt{X_S^2-2 X_S+Y_S^2+1}}\Bigg]z_L^3 +\mathcal{O}(z_L^2).
\end{align}
\end{widetext}
 The final result for the third order correction to the observed angle $\theta_O$ in terms of redshift of source and lens $z_S$, $z_L$, and observed angle $\theta_O$ is obtained by using the thin lens approximation $y_s\approx \theta_O x_S$ and the distance relation \eqref{redshift_distance} $H_0 x_S = z_S + b_1z_S^2 + b_2z_S^3 $
The full result is reported in the appendix \ref{full_bend}.

In comparison to the recent paper on dS and AdS spacetimes \cite{Takizawa22}, where the correction to third order is calculated in a strictly de Sitter background, the terms where there is coupling to the cosmological constant $\Lambda\mathcal{O}(\theta)$ are replaced by the series expansion coefficients $a_i$ and $b_i$, which, in turn, are functions of $\Omega_\Lambda$ and $\Omega_m$, from the relation \eqref{angdiam_dis}. 

One may note that for arbitrarily small $Y_S$ the corrections \eqref{full_bend} become unbounded. This is the case for Einstein Ring systems, where there is an idealized perfect alignment between source and lens. In avoiding this, we further make the approximation $y_S \approx x_L\theta_O$ motivated by the fact that the deflection happens almost completely, at order $\mathcal{O}(\alpha)$, in the lens plane \cite{Piattella_2016}, and thus $y_S \approx y_L$. This avoids any type of divergence when treating small source positions in relation to the lens.

\begin{figure}[h]
    \centering
    \includegraphics[scale=.515]{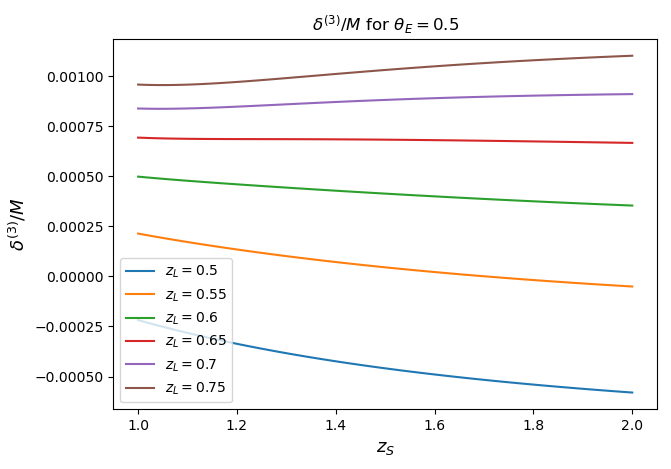}
    \caption{Plots for the ratio $\delta/M$ for different values of the observed angle $\theta_O$, with the lens at fixed redshift $z_L$.} 
    \label{fig_delta_1}
\end{figure}

In Fig.~\ref{fig_delta_1} we show the ratio $\delta/M$ as a function of $z_S$, for fixed lens redshift $z_L$ and an observed angle $\theta_O = 0.5"$ of the system. As the redshift of the lens increases, the bending angle is bigger for the same source redshift. The effect, however, gets smaller as the redshift of the lens increases. For higher source redshifts, the correction asymptotes to a constant value, as the higher order terms become increasingly small. One can also see that, for systems with a higher lens redshift, the correction changes sign, as one would expect from the Hubble flow effect of ``unbending'' the angle, as already mentioned in \cite{Piattella_2016}. 

We note that, as the lens redshift approaches the source redshift, $z_L/z_S\rightarrow 1$, the higher order correction increases in absolute value. 
This can be inferred analytically, from the form of equation \eqref{third_order_tangent}, where higher order corrections are of order $z_L^n$, but there is an overall factor $z_L/z_S < 1$.This ratio is of course strictly increasing as $z_L$ increases. For $z_S = z_L$, the bending angle has a singular behavior, as \eqref{lensdef} is ill defined. Realistically, however, cosmological strong lensing systems cannot be modeled for $z_L/z_S\approx 1$, as the thin lens approximation stops being valid.

\begin{figure}[h]
    \centering
    \includegraphics[width=1\columnwidth]{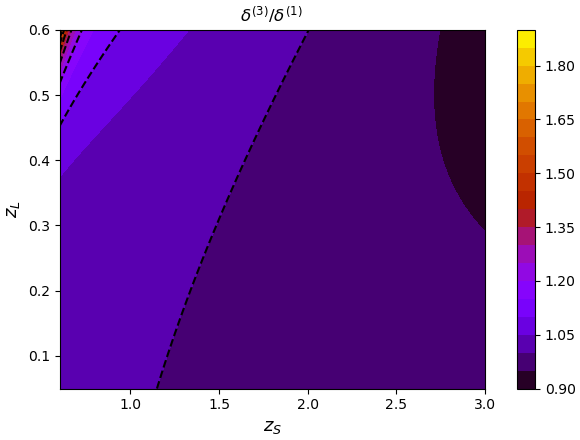}
    \caption{Ratio of the full solutions to third and first order corrections $\delta^{(3)}$ and $\delta^{(1)}$. One can see that the correction is of the same order of the first order approximation, and increases for higher values of $z_L$, as expected. The correction increases significantly as $z_S/z_L\rightarrow1$.} 
    \label{fig_delta_2}
\end{figure}

One can see from Fig.~\ref{fig_delta_2} that the correction increases significantly for higher values of $z_L$, which is expected, since the average redshift of lenses is higher than the transition redshift $z_\Lambda$ \cite{SLACS_2009}. 

It is clear from the plots that the higher order correction $\delta^{(3)}$ is at most of the same order as the first order correction found in Ref.~\cite{Piattella_2016}. Since this correction is already of order $\mathcal{O}(10^{-11})$ arcseconds for typical lensing systems, we should not expect that this new correction would be measurable by any current observation, as there are already significant systematic errors from modeling the surface density profile of the lens and its relation to the inferred mass from the luminosity \cite{SLACS_2009}.

Nonetheless, if one is to include the effect of the cosmological constant in the bending angle calculation beyond the angular diameter distance, there is a significant contribution from the cosmological behavior around the transition redshift $z_\Lambda$, which shows that the inclusion of the matter component in the calculation of the Hubble parameter becomes important for redshifts higher than this transition epoch, which is usual for current observed lensing systems.
\section{Conclusion and final remarks}
We have calculated the higher order corrections, due to the presence of matter in the cosmological model, to the bending angle $\alpha$ for a lens system, modeled as the point-like source in an expanding cosmological spacetime through the McVittie metric. Previous results using the McVittie and similar metrics \cite{Ishak_2010,mcvittie_2015} assumed a de-Sitter Universe with constant Hubble parameter $H_0$, which emulates the late-time behaviour of the Universe. To include the full matter+$\Lambda$ behavior of the late Universe, however, one must assume a non constant Hubble parameter $H(z)$, which depends also on the matter content of the Universe.

We found that the inclusion of the full matter+$\Lambda$ energy content of the Universe in the McVittie metric significantly alters the corrections to the usual bending angle assuming a Schwarzschild lens, with terms of the same order of correction in the bending angle as the one found in \cite{mcvittie_2015}, expanded to third order in the lens redshift. 

The first order correction, however is small for usual lens systems, of order $\mathcal{O}(10^{-11})$ for an average lensed Quasar system. Thus, since the previous results of \cite{Piattella_2016,Faraoni_17,Butcher_2016,Simpson_2010} already stated that the correction shouldn't influence the results for the usual lensing formalism, the result of this work strengthens this statement, showing that even beyond lensing order and with a full late-time cosmological model, the corrections to the bending angle are negligible.

\section*{Acknowledgments}
Pedro Bessa would like to thank FAPES and CAPES for the PhD scholarship, as well as CBPF for providing facilities and computational power.

\bibliography{main}
\bibliographystyle{unsrt}

\appendix
\section{Full lensing angle to 3rd order}
\label{full_bend}
For completeness, here we show the full bending angle calculation for a lens system in the thin lens approximation, $y_S\approx \theta_O x_L$.
\begin{widetext}
\begin{flalign}
    \delta^{(3)} =&\notag\\
    &\frac{\alpha}{\theta \sqrt{(z_S+b_1z_S^2 + b_2z_S^3 -1)^2 + \theta^2}} \Bigg\{1-z_S-b_1 z_s^2-b_2 z_S^3-((z_S+b_1 z_s^2+b_2 z_S^3-1)-\theta ^2)z_L \notag\\
    +&\Bigg[a_1 \Big(1+\theta^2+(\theta ^2-1) (z_S+b_1 z_s^2+b_2 z_S^3)\notag\\
    -&\theta ^2 \sqrt{(z_S+b_1 z_s^2+b_2 z_S^3-1)^2+\theta ^2} \left(-\log \left(-\sqrt{(z_S+b_1 z_s^2+b_2 z_S^3-1)^2+\theta ^2}-(z_S+b_1 z_s^2+b_2 z_S^3-1)\right)\right)\Big)\notag\\
    -&b_1 \left(b_1 z_s^2+b_2 z_S^3-\theta ^2+z_S-1\right)\Bigg]z_L^2\notag\\
    +&\Big[2a_1b_1 \Big(1+\theta^2+(\theta ^2-1) (z_S+b_1 z_s^2+b_2 z_S^3)\notag\\
    -&\theta ^2 \sqrt{(z_S+b_1 z_s^2+b_2 z_S^3-1)^2+\theta ^2} \left(-\log \left(-\sqrt{(z_S+b_1 z_s^2+b_2 z_S^3-1)^2+\theta ^2}-(z_S+b_1 z_s^2+b_2 z_S^3-1)\right)\right)\Big)\notag\\
    -&b_2 \left(b_1 z_s^2+b_2 z_S^3-\theta ^2+z_S-1\right)\notag\\
    -&a_2\Big(\theta ^2 (z_S+b_1 z_s^2+b_2 z_S^3)^2-5 \theta ^2 (z_S+b_1 z_s^2+b_2 z_S^3)+z_S+b_1 z_s^2+b_2 z_S^3+2 \theta ^4+\theta ^2-1\notag\\
    +&3\theta ^2 \sqrt{(z_S+b_1 z_s^2+b_2 z_S^3-1)^2+\theta ^2} \left(-\log \left(-\sqrt{(z_S+b_1 z_s^2+b_2 z_S^3-1)^2+\theta ^2}-(z_S+b_1 z_s^2+b_2 z_S^3-1)\right)\right)\Big)\Bigg]z_L^3\Bigg\}\Bigg|^0_{z_S},
\end{flalign}
\end{widetext}
where the constants $a_1$, $a_2$, $b_1$ and $b_2$ are given by

\begin{equation}
    a_1 = 0.225, \quad a_2 = 0.15, \quad b_1 = 0.225, \quad b_2 = 0.04875.
\end{equation}

\end{document}